\documentclass{article}% 
\usepackage{amsmath}
\usepackage{amssymb}
\usepackage{color}

\def\jyear#1{}

\jyear{2022}%

\def\al{\alpha}
\def\be{\beta}
\def\de{\delta}
\def\ga{\gamma}

\def\ep{\epsilon}

\def\te{\theta}
\def\la{\lambda}

\def\vp{\varphi}

\def\ka{\kappa}

\def\Ga{\Gamma}

\def\La{\Lambda}
\def\Om{\Omega}

 \def\calR{{\cal R}}

\def\del{\partial}
\def\na{\nabla}

\def\then{\Rightarrow}

\let\LatexFrac \frac
\def\frac[#1/#2]{\hbox{$\LatexFrac{#1}{#2}$}}
\def\Frac[#1/#2]{\LatexFrac{#1}{#2}}
\def\({\left(}
\def\){\right)}
\def\[{\left[}
\def\]{\right]}
\def\Dal{\Box}

\def\frame#1{\vbox{\hrule\hbox{\vrule\vbox{\kern2pt\hbox{\kern2pt#1\kern2pt}\kern2pt}\vrule}\hrule\kern-4pt}}

\long\def\Hide#1{}
\long\def\HideMarked#1{\hfill{$\triangleright$}}

\def\ShowLabel#1{\ref{#1}}

\def\NewSection#1{\section{#1}}
\def\NewSubSection#1{\subsection*{#1}}

\def\Acknowledgements{\section*{Acknowledgements}}

\def\ms{\medskip}

\def\eq#1{\begin{equation}#1\end{equation}}
\def\eqLabel#1#2{\begin{equation}#1\label{#2}\end{equation}}

\def\Cases#1{\begin{cases}#1\end{cases}}

\def\Align#1{\begin{aligned}#1\end{aligned}}

\def\eqs#1{\eq{\Align{#1}}}

\def\Note{\ms\begin{quote}\footnotesize\sl}
\def\EndNote{\end{quote}\normalsize}

\newtheorem{defn}{Definition}[section]
\def\Definition{\begin{defn}\leftskip=2cm}
\def\EndDefinition{\end{defn}\par}

\newtheorem{propn}{Proposition}[section]
\def\Proposition{\begin{propn}\leftskip=2cm}
\def\EndProposition{\end{propn}\par}

\newtheorem{propertyn}{Property}[section]
\def\Property{\begin{propertyn}\leftskip=2cm}
\def\EndProperty{\end{propertyn}\par}

\newtheorem{teon}{Theorem}[section]
\def\Theorem{\begin{teon}\leftskip=2cm}
\def\EndTheorem{\end{teon}\par}

\newtheorem{lemman}{Lemma}[section]
\def\Lemma{\begin{lemman}\leftskip=2cm}
\def\EndLemma{\end{lemman}\par}

\newtheorem{proofn}{Proof}[section]
\def\Proof{\begin{proofn}\leftskip=3cm\rightskip=1cm}
\def\EndProof{\end{proofn}\par\NormalStyle}

\def\Figure[#1]#2{\begin{figure}[htbp] %  figure placement: here, top, bottom, or page
   \centering
   \includegraphics[#1]{#2} }

\def\EndFigure{\end{figure}}

\def\Item[#1]{\item[#1]\leftskip=2cm}

\title{The effective Equation of State \\
in Palatini $f(\calR)$ cosmology}

\author{S.Camera$^{1,2}$, S.Capozziello$^{3,4,5}$, L.Fatibene$^{2,6}$, A.Orizzonte$^{2,6}$
\\
\\
\footnotesize
$^1$ Department of Physics, University of Torino, via P.Giuria 1, 10125  Torino (Italy)\\
\footnotesize
$^2$ Sez. Torino, Istituto Nazionale di Fisica Nucleare (INFN), via P.Giuria 1, 10125  Torino (Italy)\\
\footnotesize
$^3$ Dipartimento di Fisica ``E. Pancini'', Universit\`a  di Napoli  ``Federico II", Via Cinthia 21, 80126 Napoli (Italy)\\
\footnotesize
$^4$ Scuola Superiore Meridionale, Largo S. Marcellino 4, 80138 Napoli (Italy)\\
\footnotesize
$^5$ Sez. Napoli, Istituto Nazional di Fisica Nucleare (INFN), via Cinthia 21, 80126 Napoli (Italy)\\
\footnotesize
$^6$ Department of Mathematics, University of Torino, via C.Alberto 10, 10123 Torino (Italy)
}

\begin{document}

\maketitle

\begin{abstract} %\color{blue}
We investigate how  the cosmological Equation of State  can be used  for scrutinizing extended theories of gravity, in particular,  the Palatini $f(\cal{R})$ gravity.
Specifically, the approach consists, at first, in investigating the effective Equation of State  produced by a given model.  Then, the  inverse problem can also be considered in view of   determining  which models are compatible with a given effective Equation of State.
We consider and solve some cases and show that, for example, power-law models  are (the only models) capable of transforming  barotropic Equations of State into  effective barotropic ones. Moreover, the form of Equation of State  is preserved (only) for $f(\calR)=\calR$, as expected. In this perspective, modified Equations of State are a feature capable of distinguishing Extended Gravity with respect to General Relativity.
We  investigate also quadratic and non-homogeneous effective Equations of State showing, in particular, that  they contain the Starobinsky model and  other ones.
\end{abstract}

\NewSection{Introduction}

General Relativity  (GR) was a very successful theory from the very beginning,
soon  showing to be a self-consistent description of  gravitational interaction more precise than Newtonian gravity  at local (e.g. Solar System) scales addressing anomalies like the Mercury perihelion precession, bending of light rays by massive bodies etc. 

On the other hand, deviations  appeared very soon at bigger scales and when non-vacuum solutions were considered, mainly in clusters of galaxies  \cite{Zw1,Zw2}, 
galactic dynamics \cite{Freeman,Rubin} up to cosmological scales \cite{Trimble,Riess}. In general, these deviations require large amounts of some "dark" components  evading an explanation in the framework of the  Standard Model of Particle to be addressed at fundamental level.

In many instances, these deviations  produce  tensions between the gravitational models and observations, last but not least the one connected with the recent  Hubble parameter $H_0$ revealed comparing measurements at local and very large scales (see \cite{Eleonora} for a review). 
Generally, some authors consider  GR  essentially correct  and juggle with the matter sources to fit data (see \cite{Planck, CMB,WMAP}), 
others consider deviations as an indication that we need to modify the interaction between gravity and matter, ideally without introducing exotic sources (see, e.g.~\cite{SalFelix,MOND,Mannheim,O1,ETG1,ModBorowiec}).
Currently, we do not even reach a consensus on whether the issue is a matter of scales (since deviations appear at scales higher than galactic ones) 
or a matter of interaction between gravity and matter (since they appear in non-vacuum solutions).

However, maintaining GR as correct  leads to introduce {\it dark sources} which gravitational effects are clearly observed at astrophysical and cosmological scales although they yet eluded all attempts to be revealed at  fundamental level.
In order to fit observations,  a standard model has been obtained  where visible sources amount to  3-5\% of the observed matter-energy budget  \cite{Planck,CMB,WMAP}.

On the other hand, adopting the   {\it alternative gravity } approach, we witnessed a plethora of models which is honestly difficult to be kept under control, even challenging the realistic predictivity of these models.

In this debate, one  can deal with approaches where the same basic foundations of GR are questioned \cite{Carmen}, and then we  deal with {\it Modified Gravity}, and models where GR is recovered as a particular case on in a particular limit. In this case, we deal with {\it Extended Gravity} \cite{Rept}.

Even in specific restricted classes of models, e.g.~$f(\calR)$ gravity, one  introduces, in general,  (analytic) functions with  {\it countable infinities} of parameters.
Consequently, one would need to fix infinitely many experiments to even calibrate  models {\it before} being able to make predictions.

Usually, the situation is mitigated by (more or less arbitrarily) fixing a small subset of models, depending on a finite number of  parameters, and making predictions within this subset.
The predictions depend on the model parameters, they can be matched with observations determining best fit values  (see, e.g., \cite{Pinto,Hubbledrift,O3,O4,ExtendedCOsmologies}),
then making real predictions and occasionally even being falsified. 
However, we have no solid theoretical reasons to select  a subset rather than another class of models. 
Of course, we have indications from fundamental physics (e.g. quantum gravity) that some modifications are naturally required (see \cite{String1,O2,LQG,Loll,Carfora,CausalSets}).
However, these ones do not seem to be stringent enough yet, especially there is no consensus on the form of effective actions involving gravity.
Similarly, there are classical arguments (see, e.g., \cite{EPS,EPS1,Perlik,Distances}) to lean in one direction rather than another. Clearly, none is stringent enough to reach a general consensus, yet.

It does not really matter if we look at the current situation from the viewpoint of dark sources or  modified gravity. In any case, effective models of gravity   cannot be considered  fully satisfying  
at scales much higher than Solar System  or much smaller than laboratory ones. On the other hand, being gravity the leading interaction  from Planck to cosmological scales, 
we should be satisfied only with a theory which covers 60 orders of magnitude in distance-energy, which is unsurprisingly  hard to be achieved,
even  because many of these scales are completely observationally unexplored.

The $H_0$-tension indicates that (unless a spectacular systematic issue eventually emerges) there is something we are missing in the evolution of the universe where the same model is used from local scales up to the last scattering surface. This could be related to the Equation of State (EoS) of dark energy or some dynamical issue. In particular, there is a general consensus that anticipating the occurrence of recombination  
era would go in the direction of mitigating cosmological tensions either by some new physics \cite{Spallicci} or by improving the datasets \cite{Benetti,Vagnozzi}.

Here, we are going to consider the problem of characterizing the effective   EoS  from the  specific viewpoint of Palatini $f(\calR)$ gravity
\cite{Rept,Allemandi,Olmo,Faraoni,Faraoni2,Borowiec,Borowiec2,Wojnar}. 
The choice is due to the fact that Palatini $f(\calR)$-gravity naturally implements the {\it Ehlers-Pirani-Schild (EPS) axiomatics}, making clear how to extend GR. See \cite{EPS,EPS1,Perlik,ETG,Distances} for a discussion.

It is well known that  Palatini $f(\calR)$ gravity is equivalent, in vacuum, to Einstein gravity with a cosmological constant   \cite{Universality}.
However, assuming  a non-vacuum framework,  Palatini models  describe  conformal relations  between Einstein and Jordan frames with a prominent role played by matter.  
In these models then all is pretty similar to standard GR, exactly equal in vacuum. The only effect is that {\it real matter} is described in the Jordan frame, while
gravitational {\it effective sources} are described in the Einstein frame (and the two frames are conformally related).
In cosmology,  EoS is not preserved under conformal transformations and this fact could be relevant in selecting models consistent with observations \cite{Allemandi,ExtendedCOsmologies,Libro2}.

As we argued above,  we have very few results which claims something about the whole class of Palatini $f(\calR)$  gravity models \cite{Universality,no-go,Mana}.
Therefore, it is possible to  investigate how  constraints on $f(\calR)$ gravity can be achieved by  effective EoS.

Instead of considering the standard approach consisting in  $i)$ fixing a family of models by choosing given functions of $f(\calR)$ depending on a finite number of parameters; $ii)$ 
making predictions, and $iii)$ finally fitting observations, we try to go the other way around, looking for which models support a given effective EoS.

There are many surveys that are beginning to collect data to provide suitable dark energy EoS; see, e.g., \cite{Euclid}.
We shall show how, given this information, one can use it to determine which Palatini $f(\calR)$ gravity models produce  effective EoS. The approach  is completely different from  the one usually adopted, namely,  fixing a family of $f(\calR)$ models  and fitting the parameters vs observations. From the present point of view, one can really adopt an inverse scattering approach where underlying cosmological models are fixed by observations \cite{Elizalde}.

This gives some perspective to Extended Gravity: to some extent, we are arguing if there is a mathematical equivalence between changing source sector and the interaction. This allows to provide some information about the map which implements the equivalence \cite{Equivalence}.

The layout of the paper is the following. In Sec. 2, we outline the Palatini $f(\calR)$ gravity deriving the field equations and the main conformal properties. Sec. 3 is devoted to the main features of Palatini $f(\calR)$ cosmology while Sec. 4 is a detailed discussion of the EoS and how it can be used to select Palatini $f(\calR)$ cosmology models. Conclusions are drawn in Sec. 4.

\NewSection{Palatini $f(\calR)$ gravity}

Let us assume a spacetime $\cal M$ of dimension $\dim({\cal M})=4$.
We start from the fundamental fields $(g_{\mu\nu}, \tilde \Ga^\al_{\be\mu})$, namely a Lorentzian metric and an independent, possibly non-metric,  connection on spacetime $M$. 

An important remark is necessary at this point.
In EPS formalism,  $g_{\mu\nu}$ is a Lorentzian metric on spacetime. It comes from light rays and it defines protocols to measure distances and time lapses.
The connection $\smash{\tilde \Ga^\al_{\be\mu}}$ is independent of $g_{\mu\nu}$. It is here assumed to be torsionless describing the free fall of material points.
As such, we cannot really distinguish between two connections with the same geodesic trajectories just parameterized differently. 
Traditionally, such two connections are called {\it projectively equivalent}.
This is just because parameterizations of geodesics in spacetime are not endowed with a direct physical meaning, and we are looking for trajectories in spacetime from the very beginning.

This is also a good framework to state the Equivalence Principle  in its weakest form: {\rm there exists a single projective structure represented by a single connection describing the free fall of all forms of matter}.

Different formulations of the Equivalence Principle need to be discussed: in general, we still  can choose coordinates so that, at an event $x$, the metric $g$ is constant and the connection $\tilde \Ga$ vanishes, though not along a whole worldline, as Fermi coordinates show to be possible in  the standard GR \cite{EPS,EPS1,Perlik,ETG,Distances}.

As said, the connection $ \tilde \Ga^\al_{\be\mu}$ is not assumed to be metric {\it a priori}. It defines curvature tensors, $\smash{\tilde R^\al{}_{\be\mu\nu}}$ and its Ricci contraction $\smash{\tilde R_{\be\nu} = \tilde R^\al{}_{\be\al\nu}}$, which depend on the connection only. We can also define 
\eq{
\calR = g^{\al\be} \tilde R_{\al\be}
}
which depends on  both  he connection and the metric. It is the scalar curvature.

The dynamics is given by the Lagrangian
\eq{
L= \sqrt{g} f(\calR) + L_m[g, \Phi]
}
where   $f(\calR)$ is a regular function   
and the matter Lagrangian depends on the metric, the matter fields $\Phi$ up to their first derivatives.
It is worth noticing that
the  case $f(\calR)= \Frac[1/2\ka] \calR$ is the standard GR in Palatini formulation.

The field equations are obtained by independent variation with respect to the fields $g$ and $\Ga$
\eq{
\Cases{
f'(\calR) \tilde R_{(\mu\nu)}-\frac[1/2] f(\calR) g_{\mu\nu} = \frac[1/2] T_{\mu\nu} \cr
\tilde \na_\al \(\sqrt{g}f'(\calR)  g^{\mu\nu} \) = 0\;,
} 
}
where $T_{\mu\nu}$ is the global stress-energy tensor and $T=g^{\mu\nu} T_{\mu\nu}$ is its trace. We can contract the first equation with $g^{\mu\nu}$ to obtain the {\it master equation}
\eq{
f'(\calR) \calR -2 f(\calR) = \frac[1/2] T\,.
}
We can introduce a new conformal metric $\tilde g_{\mu\nu} = \vp \> g_{\mu\nu}$ to solve the second equation.
By fixing the conformal factor $\vp = \chi\> f'(\calR)$ for any constant $\chi$, the second equation becomes 
\eq{
\tilde \na_\al \(\sqrt{\tilde g}  \tilde g^{\mu\nu} \) = 0
\quad\iff
\quad
\tilde \Ga^{\al}_{\mu\nu} = \{\tilde g\}^{\al}_{\mu\nu}\,.
}
The function $f(\calR)$ can be chosen so that the master equation has a discrete set of zeros. For example, any analytic function, different from $f(\calR)= c\calR^2$, will do the job, although one can also consider less regular cases, still having a discrete set of zeros.
Whenever the master equation has a discrete set of zeros, one can express $\calR = \calR(T)$ as a function of matter content, in a discrete set of branches and the theory can be solved on each branch.
This allows, in field equations, solutions  for  curvature or  matter, depending on what we want to represent. 

In vacuum, for $T=0$, $\calR$ is constant and the function $f(\calR)$ can be traded with a cosmological term in field equations. This is the core of the universality theorem, saying that all Palatini $f(\calR)$ models, in vacuum, behave as the standard Einstein theory with a cosmological constant with a discrete spectrum dictated by the zeros $\calR= \calR_i(0)$ of the master equation \cite{Universality}.

In view of the second field equation,  the connection $\tilde \Ga^\al_{\mu\nu}$ is then metric on-shell, namely along solutions. Since the two metrics are conformal, we have
\eqs{
\{\tilde g\}^\al_{\mu\nu} =&  \{ g\}^\al_{\mu\nu} -\frac[1/2]\( g^{\al\ep}g_{\mu\nu} - 2 \de^\al_{(\mu} \de^\ep_{\nu)}\){}^\ast\na_\ep \ln \vp
=: \{ g\}^\al_{\mu\nu}  + K^\al_{\mu\nu}  \cr
\tilde R^{\al}{}_{\be\mu\nu} =&  R^{\al}{}_{\be\mu\nu} + \na_\mu K^\al_{\be\nu} -  \na_\nu K^\al_{\be\mu} + K^\al_{\ep\mu} K^\ep_{\be\nu} -  K^\al_{\ep\nu} K^\ep_{\be\mu}  \cr
\tilde R_{\be\nu} =&  R_{\be\nu}  -\frac[1/2] g_{\be\nu} \Dal\ln\vp -\na_{\be\nu} \ln\vp -\frac[1/2] g_{\be\nu} {}^\ast\na_\al \ln\vp {}^\ast\na^\al \ln \vp +\cr
&\quad + \frac[1/2] {}^\ast\na_\be \ln\vp {}^\ast\na_\nu \ln\vp \cr
\calR=& \vp\>  \tilde R = R -3  \Dal\ln\vp -\frac[3/2] {}^\ast\na_\al \ln\vp {}^\ast\na^\al \ln \vp
}
where ${}^\ast\na_\ep$ denotes the covariant derivative in the cases where it does not depend on any connection.
Here, covariant derivatives $\na_\al$ are with respect to the metric $g$,  the indices in
${}^\ast\na^\al \ln \vp=  g^{\al\be}\,{}^\ast\na_\be \ln \vp$
are raised by the metric $g$. The conformal factor is assumed to be positive within the region of interest $U\subset {\cal M}$ (which sets some constraint on how to fix the constant $\chi$), $\na_{\be\nu} \ln\vp = \na_\be\na_\nu \ln\vp$ is the second covariant derivative, which is symmetric on scalar 0-forms as $\ln\vp$, and $\Dal \ln\vp= g^{\be\nu}\,\na_{\be\nu} \ln\vp$ is the d'Alembert operator.

Let us stress that, as for any field equation linear in the Ricci tensor, the first field equation can be recast as follows, once one takes into account the solution of the second equation,
\eqLabel{
 \tilde R_{\mu\nu} -\frac[1/2] \tilde R \tilde g_{\mu\nu}  = \Frac[1/f'(\calR)] \( T_{\mu\nu} + \frac[1/2]f(\calR) g_{\mu\nu}\)  -\frac[1/2] \calR  g_{\mu\nu}  =: \tilde T_{\mu\nu}\;.
}{EinsteinFrame}
The relation above is the Einstein tensor equation with respect to the metric $\tilde g$ with an {\it effective stress-energy tensor} $\tilde T_{\mu\nu}$  depending on the real {\it stress-energy tensor} $T_{\mu\nu}$ and the conformal factor $\vp$. 

An important remark is in order at this point.
As a matter of fact, the physical core of Palatini $f(\calR)$ gravity, in the EPS perspective, is  the possible conformal mismatch between the metric $g$, describing distances, and the metric  $\tilde g$,   describing free fall. In Palatini $f(\calR)$ gravity, one does not assume they are the same metric as one does in standard GR, but it is left to dynamics to determine their mutual relation
(which means that also in standard GR, one does not need to assume it but it is enforced by the choice of the  Hilbert-Einstein action).

Let us comment that protocols to measure times and distances are today set in a quantum framework, which, by the way, is coherent with Special Relativity, while it is yet not known if  it is coherent with general covariance principle. 
On the other hand, the free fall of a material point is a genuinely GR issue, and nobody can  check if it is coherent with a quantum framework yet.
However, there are some  checks which  tell us that  the conformal factor does not change much within the Solar System. These tests are much harder to support experimentally at galactic or cosmological scales.

Of course, we cannot exclude that, eventually, it will turn out that standard GR is the right theory from the beginning and the conformal factor will actually be constant, that is why standard GR is regarded as a special Palatini $f(\calR)$ gravity, just for a linear $f(\calR)= \frac[1/2\ka] \calR + \La$.

Since the field equation  (\ShowLabel{EinsteinFrame}) is in the Einstein form, then the left-hand side obeys the contracted Bianchi identities. Thus, $\tilde T_{\mu\nu}$ is conserved with respect to $\tilde g$, namely $\tilde \na_\mu \tilde T^{\mu\nu}=0$, where $\tilde \na_\mu$ denotes the covariant derivative with respect to $\tilde g$ and, of course, indices of $\tilde T_{\mu\nu}$ are raised by $\tilde g$.

Because of this form of the field equations, using the metric $\tilde g$ is  called the {\it Einstein frame(work)}, while using the metric $g$ is  called the {\it Jordan frame}.
Endless discussions have been spent to discuss whether the true physical metric on $\cal M$ is, $g$ or $\tilde g$, regarding the other one as a mathematical trick \cite{Allemandi}.

It is worth noticing that,
in EPS formulation, there is no reason to believe that one should do everything with one metric only. As a matter of fact, EPS point of view is  that quantum mechanics and the local non-gravitational metric $g$ may be conformal but not equal to the metric $\tilde g$, which describe the free fall of macroscopic material points.

Of course, none of these interpretations is relevant to the formulae we write down here; they just explain the names we use to denotes objects. 
When we call $\tilde T_{\mu\nu}$ the {\it effective stress-energy tensor}, we exactly mean that it encodes gravitational effects, not quantum effects.
If we set up quantum experiments to test matter composition, we expect to see $T_{\mu\nu}$, not $\tilde T_{\mu\nu}$. If it can sound strange that we never realized the conformal mismatch between the two frameworks, actually this is precisely the case since we are not able to resolve GR at  quantum level, yet.  In fact,  only Newtonian gravity can be successfully tested on quantum systems \cite{BoseGravity}. This goals, at the moment, is quite far to be achieved for GR and the other relativistic theories of gravity \cite{Altschul}.

If it is there, the discrepancy will sooner or later appear clearly. Nowadays one can reasonably expect to see effects at galactic or cosmological scales, which are, by the way, exactly the scales at which dark sources become relevant while they are not yet detected at laboratory scales.

By the way, arguing that Einstein frame is characterized by Einstein-like field equations is just wrong. Of course, field equations can be {\it also} recast as Einstein equations for $g$; namely as
\eqs{
 R_{\be\nu}   -&\frac[1/2] R g_{\be\nu}
= \Frac[1/2 f'(\calR)  ] \( T_{\be\nu} +  f(\calR) g_{\be\nu} \)
+\frac[1/2] g_{\be\nu} \Dal\ln\vp + \na_{\be\nu} \ln\vp +\cr
&+\frac[1/2] g_{\be\nu} \na_\al \ln\vp \na^\al \ln \vp - \frac[1/2] \na_\be \ln\vp \na_\nu \ln\vp 
 -\frac[1/2] R g_{\be\nu} =:\hat T_{\be\nu}\;.
}
Also, in this case, $\hat T_{\be\nu}$ is called the {\it Jordan stress-energy tensor}, it is conserved with respect to the metric $g$, namely $\na_\be \hat T^{\be\nu}=0$ (indices raised by $g$, here).
We shall not use this below, but it just shows that it is not true that field equations in Jordan frame are not in Einstein form. Distinguishing Jordan and Einstein frames needs to be done at the level of action functionals, not at the level of field equations \cite{Magnano, ETG3}. 
  
One can  finally show that the real stress-energy tensor $T_{\mu\nu}$ is conserved with respect to $g$, namely $\na_\mu T^{\mu\nu}=0$ since it comes as a variational derivative of a covariant matter Lagrangian $L_m[g, \Phi]$. 

Although one can show this directly \cite{Ferraris}, we can also have a direct hint by the fact that $T_{\mu\nu}$ does not  depend on the gravitational Lagrangian we couple  to it. Clearly $T_{\mu\nu}$ is conserved in standard GR.

\NewSection{Palatini $f(\calR)$ cosmology}

After the above discussion on  Palatini $f(\calR)$ gravity,  we want to apply it to cosmology, namely to  spatially flat, homogeneous and isotropic Friedman-Le ma\^\i tre-Robertson Walker (FLRW) cosmology.
Let us  define 
\eq{
g= -c^2 dt^2 +a^2(t)\,\(\Frac[dr^2/1-k\,r^2] + r^2\,\(d\te^2 + \sin^2\te\,d\phi^2\)\)\;,
}
where $g$ is the FLRW metric, defined after the line element $ds^2=g_{\mu\nu}dx^{\mu}dx^{\nu}$.
We choose to normalize the {\it scale factor} $a_o= a(t_o) = 1$ to unity today at time $t_o$. 
In this normalization, the scale factor $a(t)$ is assumed  dimensionless with $[k]= L^{-2}$, and $[r]=[ct]=L$.
It is worth saying that   here we have two metrics and, in principle, we should declare which one is required to be spatially homogeneous and isotropic.

In view of the master equation, we can see that the conformal factor $\vp$ is a function of time $t$ only.
Suppose $g$ is in FLRW form in coordinates $(t, r, \te, \phi)$, then we have
\eqs{
\tilde g = \vp\> g=& -c^2\vp \>dt^2 + \vp \>a^2(t)\(\Frac[dr^2/1-kr^2] + r^2\(d\te^2 + \sin^2(\te) d\phi^2\)\)=\cr
=& -c^2 d\tilde t^2 + \tilde a^2(\tilde t)\(\Frac[dr^2/1-kr^2] + r^2\(d\te^2 + \sin^2(\te) d\phi^2\)\)
}
where we set $d\tilde t = \sqrt{\vp(t)} \> dt$ and $\tilde a(\tilde t)= \sqrt{\vp(t)}\> a(t)$. 
Hence $\tilde g$ is in FLRW form in coordinates $(\tilde t, r, \te, \phi)$, therefore it is spatially homogeneous and isotropic.

If we restrict to a conformal factor such that $\vp(t_o)=1$, the conformal transformation preserves the normalization of the scale factor, namely also $\tilde a(\>\tilde t_o)=1$.

The second step is that being the metric in FLRW form, it  means that the corresponding Einstein tensor, as well as the Jordan tensor 
\eq{J_{\mu\nu}=f'(\calR) \tilde R_{\mu\nu}-\frac[1/2] f(\calR)g_{\mu\nu},}
is a {\it perfect fluid stress-energy tensor}. In other words, there exist two functions of time, $\rho(t)$ and $p(t)$, and a future-directed timelike unit-vector $u^\mu$ such that
\eq{
T_{\mu\nu} = c^2(\rho+p)\> u_\mu u_\nu + p \>g_{\mu\nu}
\qquad
\tilde T_{\mu\nu} = c^2(\tilde \rho+\tilde p)\> \tilde u_\mu \tilde u_\nu + \tilde p\> \tilde g_{\mu\nu}
}
where $\rho$ and $p$ are called the {\it energy-density} and {\it pressure} of the fluid.
On the other hand,   $\tilde \rho$ and $\tilde p$ are  the {\it effective} energy-density and pressure.
Immediately we have $ u^\mu = \sqrt{\vp} \>\tilde u^\mu$ and $\tilde u_\mu = \sqrt{\vp}\>  u_\mu$, so that $g(u, u)=-1=\tilde g(\tilde u, \tilde u)$.

In view of the relation (\ShowLabel{EinsteinFrame}) between stress-energy tensors, we have
\eqs{
\tilde T_{\mu\nu} =&  c^2\vp\>(\tilde \rho+\tilde p)\>  u_\mu u_\nu + \vp\>\tilde p\> g_{\mu\nu}=\cr
=&  \frac[ c^2/f'] (\rho+p)\> u_\mu u_\nu + \(\frac[p/f'] + \frac[f/ 2f']  -\frac[1/2] \calR  \) g_{\mu\nu} 
}
from which, we get  the relations between real and effective sources, namely
\eq{
\chi (f')^2 \>\tilde p = p + \frac[1/ 2]  \( f-f' \calR\)
\qquad
\chi (f')^2 \tilde \rho =    \rho -  \frac[1/ 2]  \( f-f' \calR\)
}
On each branch of the theory, we have $\calR=\calR(T)$ and $T=3p-\rho$, which expresses also $f$ and $f'$ as a function of $p$ and $\rho$.
In other words we have 
\eq{
\tilde \rho = \tilde \rho(\rho, p)
\qquad
\tilde p = \tilde p (\rho, p)
}
As we supply the EoS, $p=p(\rho)$, for real matter, this is a parameterized form of EoS, $\tilde p=\tilde p(\tilde \rho)$, for effective matter.

Let us stress that the change of EoS, when changing from Jordan to Einstein  frame, can be considered as the only effect introduced by the Palatini $f(\calR)$ cosmology. It is worth discussing how this change can occur and how it can put constraints on related models to be matched with observations.

\NewSection{Selecting Palatini $f(\calR)$ cosmologies from EoS}

 Suppose  we measure $\tilde H(\tilde z)$ that is equivalent, according to standard cosmology,  to know the (total) effective density $\tilde \rho(\tilde a)$.
 Then, by conservation, we have $\tilde p= -\(\tilde a\> \tilde \rho'  +\frac[1/3]\tilde\rho\)$.
 Once we have $\( \tilde\rho(\tilde a ), \tilde p(\tilde a )\)$, that is equivalent to have the effective EoS in the implicit form $\tilde F(\tilde \rho, \tilde p)=0$,
  we can investigate which $f(\calR)$ models reproduce the observed effective behaviour. See also \cite{Elizalde} for a general discussion.
 
 For example,  it is possible to show that deviations from the barotropic EoS are  signals of  dynamics departing from power-law models. On the other hand, if we have a quadratic EoS,  this is strong constraint on possible dynamics one can consider. In other words,   measuring $\tilde H(\tilde z)$ gives  a lot of information on the sources related to the EoS.
 
Let us start by reviewing some  particular models. After we shall consider the inverse problem, i.e. how to constrain $f(\calR)$ cosmologies by real and effective EoS.

\NewSubSection{The Starobinsky model}

Let us start from the  Starobinsky model, namely $f(\calR)= \frac[1/2\ka]\( \calR + \ep \calR^2\)$.
The master equation in this case is 
\eq{
-\calR=\calR -2 \calR  = \ka T= \ka (3p-\rho)
\quad\then
\calR= \ka (\rho-3p)
}
and consequently
\eq{
f =  \frac[ \ka(\rho- 3p) /2\ka]\(   1 +\ka \ep   (\rho- 3p)\)
\qquad
f' =  \frac[1/2\ka]  \( 1+ 2\ka\ep   (\rho- 3p)\)
%\quad\thenf-f'\calR=   -\frac[\ka\ep/2](\rho- 3p)^2
}
 If the real matter is barotropic, namely the real EoS is $p=w\rho$, we have the effective sources as
\eqLabel{
\chi (f')^2 \>\tilde p = w\rho - \frac[\ka\ep/ 4]    (1-3w)^2\rho^2
\qquad
\chi (f')^2 \tilde \rho =    \rho +\frac[\ka\ep/ 4]   (1-3w)^2\rho^2
}{EoSParametrizzate}
These are already effective EoS in a parameterized form, with $\rho$ as a parameter.
This already shows that effective matter is not even barotropic, in fact we have
\eq{
\tilde p  = \Frac[ 4w - \ka\ep    (1-3w)^2\rho /  4 +\ka\ep   (1-3w)^2\rho]\> \tilde \rho=\tilde w(\rho) \> \tilde \rho
} 
We see that the effective EoS is not barotropic since $\tilde w$ is not constant (unless one expands around a configuration or $w=\frac[1/3]$). Albeit, in a sense, this is an  exact EoS to the extent in which the real EoS is assumed to be exact.  From an observational point of view, the dark energy EoS $w_\mathrm{DE}(z)$, which is aimed to be measured by several cosmological surveys, cannot be exactly barotropic if we want to describe the evolution of cosmic fluid where the global EoS is not always strictly constant. 

We can also eliminate the parameter $\rho$ to have the effective EoS in implicit form as $\tilde F(\tilde \rho, \tilde p)=0$. In fact, starting from (\ShowLabel{EoSParametrizzate}), we can divide and solve for $\rho$ 
\eq{
\tilde p  = \Frac[ 4w - \ka\ep    (1-3w)^2\rho /  4 +\ka\ep   (1-3w)^2\rho]\> \tilde \rho
\qquad\then
\rho  =\Frac[4 / \ka\ep(1-3w)^2] \Frac[w \tilde \rho - \tilde p / \tilde p+\tilde \rho] \;.
}
Summing and eliminating $\rho$, we have
\eq{
  \(  (1 +5w )\tilde \rho-(7+ 3w  )\tilde p \)^2 =\frac[16 \ka / \chi\ep ] (1+w) (w \tilde \rho - \tilde p )\;, 
}
which is  the effective EoS.

 \NewSubSection{Power-law models}
 
 Let us discuss now  power-law models,  $f(\calR)= \frac[1/2\ka] \calR^n$, with $n\not=2$.
Also in this case, the master equation  is particularly simple
\eq{
(n-2)\calR^n=n  \calR^n -2  \calR^n = \ka T= \ka (3p-\rho)\;,
\quad\then
\calR^n= \frac[\ka/2-n] (\rho-3p)
}
and consequently
\eq{
f =   \frac[1/2(2-n)] (\rho-3p)\,,
\qquad\qquad
f' =   \frac[n/2\ka] \(\frac[\ka/2-n]\)^{\frac[n-1/n]} \(\rho-3p\)^{\frac[n-1/n]}\;.
%\quad\then f-f'\calR =    \frac[1-n /4-2n] (\rho-3p)
}
 If the real matter is barotropic, namely the real EoS is $p=w\rho$, we have effective sources as
\eqs{
\chi (f')^2 \>\tilde p =& \frac[ 5w +1  -wn - n / 8-4n] \rho
\qquad
\chi (f')^2 \tilde \rho =   \frac[9-3n -3wn -3w / 8-4n] \rho
\cr
&\tilde p  = \Frac[ 5w +1  -wn - n /  9-3n -3wn -3w]\> \tilde \rho=\tilde w \> \tilde \rho\;.
}
This means that, in the case of power-law models, a  barotropic EoS is mapped in a barotropic one and vice versa.

\NewSubSection{The inverse problem}

Suppose we want to go the other way around, that is we measure an EoS  and we want to reconstruct the related  $f(\calR)$ cosmological model.

For the sake of simplicity,   let us first consider a barotropic real EoS, $p=w\>\rho$, and a different barotropic effective EoS, $\tilde p = \tilde w\> \tilde \rho$.
By specializing to this case we have the relation between real and effective sources as well as the master equation
\eq{
\Cases{
\chi (f')^2 \>\tilde w \> \tilde\rho  = w\>\rho + \frac[1/ 2]  \( f-f' \calR\)			\cr
\chi (f')^2 \tilde \rho =    \rho -  \frac[1/ 2]  \( f-f' \calR\)		\cr
f' \calR -2 f = \frac[1/2] (3w-1)\rho
}
\quad\then
  \Frac[(3\tilde w-1)(w+1)/3w\tilde w-5w +7 \tilde w- 1 ] \Frac[ f'/f] 	 =    \Frac[1/\calR ]
}
which integrates to
\eq{
f(\calR)	 =      c \calR^{ \frac[3w\tilde w-5w +7 \tilde w- 1 / (3\tilde w-1)(w+1) ]} =: c\calR^n\;.
}
This result shows that models with $f(\calR)= c\calR^n$ send real barotropic sources in (different) effective barotropic sources.
If one wants to preserve the barotropic index, namely having $\tilde w=w$ that is possible if and only if 
\eq{
 \frac[3w^2+2w- 1 / 3 w^2  + 2 w-1 ]=n=1\;.
 }
 Therefore, standard GR is the only model in which effective and real EoS coincide.
 
 Furthermore, it is worth noticing   that $w=\frac[1/3] =\tilde w$, as well as $w=-1=\tilde w$, are solutions for any $n$ so that, given both EoS, the function $f(\calR)$ is not, in general, uniquely determined by a single correspondence of EoS, but rather by the map which transforms any real EoS in the corresponding effective EoS.

If we had a reason to believe in a specific model, the direct problem would be fine. 
It makes predictions about the gravitational sector which may be confirmed or disproved.
But currently we do not. 
If we do not consider dark sources, we have no (or very little) reasons to go beyond the standard GR.
Also considering dark sources, one can go for modifying the matter or the gravitational sector. 
Preferring to modify the gravitational sector,  the motivation remains purely phenomenological: one should set up a model capable of reproducing an observed  phenomenology.

Solving the equation for $f(\calR)$, in general, is not easy but it gives us a different information about the model rather than the direct problem. However, there are more cases in which it can be done.

\NewSubSection{The linear effective EoS}

Let us consider an effective EoS like
\eqLabel{
\al\>\tilde p + \be\> \tilde \rho = \ga
%\qquad
%\tilde p= \al\> \tilde \rho^2 + \be\>\tilde \rho
}{LinearEffectiveEoS}
which is defined up to a  factor that we can fix  without changing the EoS itself.
Thus with no loss of generality, we can set $\ga=1$ to parameterized all non-barotropic cases.
This is the simplest modification of barotropic EoS and it allows non-zero pressure in vacuum.
We also assume barotropic real EoS, i.e.\ $p=w\rho$.
Then we have
\eq{
\Cases{
f' \calR -2f =  \frac[1/2](3w-1)\rho
\quad\then
\rho = \frac[2/3w-1] f' \calR -\frac[4/3w-1] f			\cr
\chi (f')^2 \al\>\tilde p =  \frac[\al(w+1)/ 2(3w-1)] f' \calR	 - \frac[\al(5w+1)/ 2(3w-1)]   f			\cr
\chi (f')^2 \be\>\tilde \rho =    \frac[3\be(w+1)/ 2(3w-1)]  f' \calR 	 - \frac[\be(3w+7)/ 2(3w-1)]   f 
}
}
from which we obtain
\eqLabel{
 2\chi (3w-1)  (f')^2 =  \( \al +    3\be\)(w+1) f' \calR 	 - \( \al(5w+1)+ \be(3w+7)\)   f \,.
}{LinearEoSCond}
We can now perform the transformation $f(\calR)= \La + \al \calR + \calR^2 F(\calR)$ (hence $f' =  \al + 2\calR F + \calR^2 F'$)  so that the condition becomes
\eqs{
&2\chi (3w-1)  (F')^2\calR^4 
+\(8\chi (3w-1)  F  -(w+1) (\al + 3\be)\) F'\calR^3+\cr
&\quad
+ \(\chi(8F^2 + 4c F') + (\al - \be)F\)(3w - 1)\calR^2
 +\cr 
&\quad
+ \(8\chi (3w-1)F  + 4\al w + 4\be\) c\calR +\cr 
&\quad
+\((5\al + 3\be)w + \al + 7\be)\La + 2\chi c^2 (3 w - 1) \)=0
}
Assuming the analytic function $F(\calR)= a_0+ a_1\calR+\frac[1/2]a_2 \calR^2+\dots$ and for generic values of the parameters, we get
\eq{
\La=  \Frac[-2\chi c^2(3 w - 1) / (5 w  + 1)\al + (3 w+ 7)\be]
\qquad
a_0  =- \Frac[\al w +\be/  2\chi (3w-1) ]
\qquad
\dots
}
giving rise to a recurrence formula to compute the following coefficients from the previous ones.
Whether this formal series truncates to a polynomial depends on specific values of $\al, \be, w$.

\NewSubSection{The quadratic effective EoS}

Let us now consider an effective EoS like
\eqLabel{
(\al\>\tilde p + \be\> \tilde \rho)^2 = \la \tilde p+ \ga\>\tilde \rho\,.
%\qquad
%\tilde p= \al\> \tilde \rho^2 + \be\>\tilde \rho
}{EffectiveEoS}
As a special case,  we get the Starobinsky model for
\eq{
\al=-3w-7
\qquad
\be=5w+1
\qquad
\la=  - \Frac[16 \ka (1+w) /  \chi\ep]
\qquad
\ga= \Frac[16 \ka (1+w) w/  \chi\ep]\,.
}
%\eq{ \(  (1 +5w )\tilde \rho-(7+ 3w  )\tilde p \)^2 =\frac[16 \ka (1+w) w/  \chi\ep]  \tilde \rho - \frac[16 \ka (1+w) /  \chi\ep]\tilde p }
Let us change coordinates $(\tilde \rho, \tilde p)\mapsto (\tilde e, \tilde p)$ by setting  $\tilde e=\al\>\tilde p + \be\> \tilde \rho$ 
(hence $\tilde \rho=  \frac[1/\be] \tilde e-\frac[\al/\be]\>\tilde p $)
so that the effective EoS becomes
\eqLabel{
\be \tilde e^2 = \be \la \tilde p+  \ga \tilde e-\al\ga\>\tilde p 
\quad\then
( \be\la -\ga\al )\>\tilde p = \be \tilde e^2 - \ga \tilde e\;.
}{SimpEffectiveEoS}
Then we can consider the equations for the new pressure and density together the master equation and use these to write a differential equation for the function $f(\calR)$ alone.
We proceed as above and, from the master equation, we get
\eq{
f' \calR -2 f = \frac[1/2] (3w-1)\rho
\quad\then
\rho=  \frac[2/ 3w-1]f' \calR -  \frac[4/ 3w-1] f
}
Then we make a linear combination of the equations for the effective sources  obtaining
\eqs{
\chi (f')^2 \>\tilde e =& \(\al w +  \be \) \rho +  \frac[\al-\be/ 2]  \( f-f' \calR\)=\cr
 =&   \frac[(\al+ 3\be) (w+1)  / 2(3w-1)] f' \calR
-  \frac[\al(5 w +1)+\be(3 w+7)/ 2(3w-1)]   f  
}
%check\eq{  \frac[-8\al w+ 3w \al  -\al -3w\be -8 \be +\be/ 2(3w-1)] }
while from the equation for the effective pressure, using the effective EoS in the form (\ShowLabel{SimpEffectiveEoS}), we have
\eqs{
\chi (f')^2 ( \be \tilde e^2 - \ga \tilde e ) =&  \frac[2w( \be\la -\ga\al )/ 3w-1]\( f' \calR -2 f \)  + \frac[ \be\la -\ga\al / 2]  \( f-f' \calR\)=\cr
=&  \frac[   (\be\la-\ga\al)(w+1)  / 2(3w-1)]  f' \calR
 + \frac[ (\ga\al -\be\la)(5w+1)  / 2(3w-1)]  f\,,
}
from which we get
\eqs{
 \chi (f')^2   \tilde e^2 =  
 \frac[   ( 3\ga+ \la)(w+1)  / 2(3w-1)]  f' \calR
-  \frac[  \la(5w+1)+\ga(3 w+7)/ 2(3w-1)]   f\,,   
}
and then
\eqs{
&\(\frac[(\al+ 3\be) (w+1)  / 2(3w-1)] f' \calR-  \frac[\al(5 w +1)+\be(3 w+7)/ 2(3w-1)]   f  	\)^2 
= \cr
&\quad
=\chi (f')^2 \>\(  \frac[   ( 3\ga+ \la)(w+1)  / 2(3w-1)]  f' \calR-  \frac[  \la(5w+1)+\ga(3 w+7)/ 2(3w-1)]   f   \)\,.
} 
 The last equation is non-linear and already quite difficult to study, although the effective EoS we selected is not very complicated. 
However, it contains the information we are looking for, being a differential equation for the function $f(\calR)$.
It provides us with all possible Palatini $f(\calR)$ models reproducing the effective EoS we started from.
It is no surprising that among solutions we find $f(\calR)= a_1\calR + a_2 \calR^2$.  %\hfill$f' \calR= a_1\calR + 2a_2\calR^2$
In general, if we expand the equation as a polynomial in $\calR$, it is possible to get equations to link $(a_1, a_2, \al,\be, \ga, \la, w)$. 

What we would like to do now is providing EoS parameters $\al, \be, \ga, \la, w$ and computing $a_1, a_2$ to determine $f$.
However, in this case, we already see that we have a 4-degree  polynomial (with no constant term), thus we get 4 equations.
With 2 unknowns, we have little hope to determine them uniquely.
In fact we can define $\ep=\ep(\al,\be,\ga,\la, w)$ but we are left with 3 equations from which we cannot solve for $\ka$, which is left undetermined, i.e. they are constraints for $\al,\be,\ga,\la, w$.

However, in this case we can try and solve for $\al, \be, \ga, \la$ and we obtain 
\eq{
\al= -(3w+7)c
\quad
\be=(5w+1)c
\quad
\ga=\frac[16\ka/\chi\ep]w(w+1)c^2
\quad
\la=-\frac[16\ka/\chi\ep](w+1)c^2
}
where we set $a_1=\frac[1/2\ka]$ and $a_2=\ep$.
That is exactly the  Starobinsky model for any value of $c$.

However, this shows precisely that Starobinsky models account for a 1-parameter family of models with effective EoS as
\eq{
(\al\>\tilde p + \be\> \tilde \rho)^2 = \la \tilde p+ \ga\>\tilde \rho
}
which, however, are a 3-parameters family. Thus we expect other models producing the same type of effective EoS.

In fact, Starobinsky model is not the only model producing effective EoS in the form (\ShowLabel{EffectiveEoS}). 
As a matter of fact, if we assume a function 
\eq{
f(\calR)= \Frac[1/2\ka] \Frac[(\calR+3\La)^3/27\La^2]
}
and we repeat the above argument, we see that the
effective EoS is in the form (\ShowLabel{EffectiveEoS}) with
\eq{
\al= -3c
\qquad
\be= c
\qquad
\ga= -\Frac[8\ka(w-1)\La / 3\chi(w+1)]c^2
\qquad
\la= -\Frac[8\ka(3w+1)\La / 3\chi(w+1)]c^2
}
Then we do not get more up to a polynomial of degree 10.

Finally, we also have 
\eq{
f(\calR) = \frac[1/2\ka] \( \La + \calR\)
}
with effective EoS
\eq{
\al= c
\qquad
\be= -wc
\qquad
\ga= -\Frac[\ka w(w+1)\La / \chi]c^2
\qquad
\la= \Frac[\ka(w+1)\La / \chi]c^2\,.
}
However, even if this is in the form (\ShowLabel{EffectiveEoS}), one can simplify it to a linear effective EoS as
\eq{
\tilde p - w \tilde \rho= \Frac[\ka(w+1)\La / \chi] 
}
although generically not homogeneous as a barotropic law.
The same happens for the model with
\eq{
f(\calR) = \frac[1/2\ka] \( 2\La + \calR\)^2\,,
}
which is clearly a generalisation of the Starobinsky one.

 \NewSection{Conclusions and perspectives}
 Gravitational theories have a notoriously tricky relation with observations. It is well known that, by defining observables, as generally invariant quantities
purely metric  covariant theories, in vacuum, has no observables at all  \cite{Rovelli,Utiyama,Bruzzo,Janyska,Holeargument}.
Gravitational observables arise only when matter is adopted to probe theories. Specifically, 
in cosmology,  we need to understand how observable quantities relate to the theoretical framework, like, for example, that given by the
Cosmological Principle defining some preferred observers. Certainly, in cosmology, we have many constraints on what we {\it can} measure, making it a coarse grained approach.

In this perspective,  surveys like  \textit{Euclid}  can provide information about the (effective) EoS of dark sources.
Thus one should investigate what information we can achieve  within  theories of gravity like Extended Gravity or Modified Gravity.

 Here, we showed that, with the exception of standard GR, any Palatini $f(\calR)$ model comes with a non-trivial map between real and effective EoS,
in the sense that  standard GR is the only theory which preserves EoS.

If we assume barotopic real EoS, power-law  models are the only models in which effective EoS are also barotropic, with the exception of $f(\calR)= c\calR^2$ which is degenerate
since, for it, the master equation is identically satisfied.

When we take non-barotropic effective EoS,  we can also  study which models are allowed. 
This is done by starting from the definition of effective pressure and density together with the master equation and turn this into a (non-linear) differential equation for the model function $f(\calR)$.
This function is not easy to be solved even in the simplest cases.
However, if we expand in series of $f(\calR)$,  we can compute   approximations of general  solutions. 

Generically, we can solve the conditions for $f'$. 
For example, we can solve Eq.(\ShowLabel{LinearEoSCond}) on two different branches to obtain $f'= A(\calR, f)$ which, together with the trivial equation $\calR'=1$,
can be considered as a solution of the dynamical system defined by the vector field 
\eq{
X= \Frac[\del/\del \calR] +  A(\calR, f) \Frac[\del/\del f] 
}
which shows that, at least locally, one has solutions for  $f(\calR)$.

Here we considered  simple toy models with a single real barotropic fluid and still we obtained effective EoS which are non-trivial.
Future investigations have to be devoted to extend to general real EoS and mixtures.
Having non-linear EoS is not really difficult to be dealt with. See, for example,  \cite{Pinto, Hubbledrift}.
After all, EoS are themselves an approximation of matter equations, which are certainly valid in a regime and extending it out of this domain is problematic.
For example, ordinary matter is described as dust today since one can neglect non-gravitational interactions among far away galaxies, 
but of course such an assumption is at least questionable in the early universe.

Moreover, we already showed that when one allows mixtures, the cosmic evolution is determined by the {\it total abundance} $\Om(a)$.
However, there are different mixtures producing the same function $\Om(a)$.
Still we need to define results which hold for any $f(\calR)$ to realize  if Palatini approach  can be physically sound.

Let us consider the conjecture stating that given an EoS, one can find $f(\calR)$ which produce it.
We already proved that different models can sends a single real EoS to a single effective EoS.
We still have no evidences that two correspondences between real and effective EoS cannot be reproduced by different models.
If this is the case, namely, we can have different models which cannot be distinguished in cosmology, which is very interesting.

If there are specific effective EoS that cannot be produced at all would be also very interesting since it would allow to disprove all Palatini $f(\calR)$ models at a time.
Finally, as we said, if there is a one-to-one correspondence between $f(\calR)$ and a map between real and effective EoS
 would be a proof which could be tested, in principle, by observations.
 Clearly the approach can work also for other theories of gravity like teleparallel or symmetric teleparallel ones \cite{Carmen,Cai}. This will the the topic of forthcoming studies.
 
\Acknowledgements
This paper is based upon work from COST Action CA21136 {\it Addressing
observational tensions in cosmology with systematics and fundamental
physics} (CosmoVerse) supported by COST (European Cooperation in Science and
Technology).  It is also supported by INdAM-GNFM.
We  acknowledge the contribution of Istituto Nazionale di Fisica Nucleare  ({\it CSN4 iniziativa specifica} QGSKY and {\it CSN2 esperimento} Euclid), the local research project {\it Metodi Geometrici in Fisica Matematica e Applicazioni} (2021) of Dipartimento di Matematica of University of Torino (Italy).  
LF acknowledges the Department of Applied Mathematics of University of Waterloo (ON-Canada) for the invitation as a visitor researcher and the financial support  for the sabbatical leave.


\end{document}